\documentclass[prb,twocolumn,superscriptaddress,showpacs]{revtex4}
\usepackage{epsfig}
\begin{document} 
\newcommand{\dprime}{{\prime\prime}}
\newcommand{\be}{\begin{equation}}
\newcommand{\den}{\overline{n}} 
\newcommand{\ee}{\end{equation}}
\newcommand{\bea}{\begin{eqnarray}} 
\newcommand{\eea}{\end{eqnarray}}
\newcommand{\nn}{\nonumber} 
\newcommand{\vk}{{\bf k}}
\newcommand{\vN}{{\bf \nabla}}
\newcommand{\vA}{{\bf A}}
\newcommand{\vE}{{\bf E}}
\newcommand{\vj}{{\bf j}}
\newcommand{\vJ}{{\bf J}}
\newcommand{\vs}{{\bf v}_s}
\newcommand{\vn}{{\bf v}_n}
\newcommand{\vv}{{\bf v}} 
\newcommand{\la}{\langle}
\newcommand{\ra}{\rangle} 
\newcommand{\ph}{\phi} 
\newcommand{\dg}{\dagger}
\renewcommand{\vr}{{\bf{r}}} 
\newcommand{\vq}{{\bf{q}}}
\newcommand{\vQ}{{\bf{Q}}} 
\newcommand{\hj}{\hat{\alpha}}
\newcommand{\hx}{\hat{\bf x}} 
\newcommand{\hy}{\hat{\bf y}}
\newcommand{\hz}{\hat{\bf z}}
\newcommand{\vS}{{\bf S}} 
\newcommand{\cV}{{\cal U}}
\newcommand{\cD}{{\cal D}} 
\newcommand{\tnh}{{\rm tanh}}
\newcommand{\sh}{{\rm sech}} 
\newcommand{\vR}{{\bf R}}
\newcommand{\crx}{c^\dg(\vr)c(\vr+\hx)}
\newcommand{\crkubox}{c^\dg(\vr)c(\vr+\hat{x})}
\newcommand{\pll}{\parallel} 
\newcommand{\crj}{c^\dg(\vr)c(\vr+\hj)}
\newcommand{\crmj}{c^\dg(\vr)c(\vr - \hj)}
\newcommand{\sumall}{\sum_{\vr}} 
\newcommand{\sumx}{\sum_{r_1}}
\newcommand{\nabj}{\nabla_\alpha \theta(\vr)} 
\newcommand{\nabx}{\nabla_1\theta(\vr)} 
\newcommand{\sumy}{\sum_{r_2,\ldots,r_d}}
\newcommand{\krj}{K(\vr,\vr+\hj)} 
\newcommand{\sigr}{|\psi_0\rangle}
\newcommand{\sigl}{\langle\psi_0 |}
\newcommand{\sier}{|\psi_{\Phi}\rangle}
\newcommand{\siel}{\langle\psi_{\Phi}|}
\newcommand{\sumrj}{\sum_{\vr,\alpha=1\ldots d}}
\newcommand{\krw}{K(\vr,\vr+\hx)} 
\newcommand{\Dtheta}{\Delta\theta}
\newcommand{\rhonew}{\hat{\rho}(\Phi)}
\newcommand{\rhoold}{\hat{\rho_0}(\Phi)} 
\newcommand{\dt}{\delta\tau}
\newcommand{\cP}{{\cal P}} 
\newcommand{\cS}{{\cal S}}
\newcommand{\vm}{{\bf m}} 
\newcommand{\hnr}{\hat{n}({\vr})}
\newcommand{\hnm}{\hat{n}({\vm})} 
\newcommand{\del}{\hat{\delta}}
\newcommand{\upa}{\uparrow} 
\newcommand{\dna}{\downarrow}
\newcommand{\dnk}{\delta n_{\vk}}
\newcommand{\dnks}{\delta n_{\vk,\sigma}}
\newcommand{\dnkp}{\delta n_{\vk '}}

\title{Fermi liquid interactions and the superfluid density in d-wave
superconductors}
\author{Arun Paramekanti}
\affiliation{
Tata Institute of Fundamental Research, Mumbai 400 005, India}
\affiliation{
Department of Physics and Kavli Institute for Theoretical Physics, University 
of California, Santa Barbara, California 93106--4030}
\author{Mohit Randeria}
\affiliation{
Tata Institute of Fundamental Research, Mumbai 400 005, India}

\begin{abstract}
\vspace{0.1cm}
We construct a phenomenological superfluid Fermi liquid theory for a
two-dimensional d-wave superconductor on a square lattice,
and study the effect of quasiparticle interactions on the superfluid density. 
Using simple models for the dispersion and the Landau interaction function, 
we illustrate the deviation of these results from those for the isotropic 
superfluid. This allows us to reconcile the value and doping dependence of 
the superfluid density slope at low temperature obtained from penetration 
depth measurements, with photoemission data on nodal quasiparticles.

\typeout{polish abstract}
\end{abstract}

\maketitle

\noindent

The high temperature superconductors appear to support well-defined
quasiparticle (QP) excitations at low temperatures ($T\!\!\ll\!\!T_c$) as 
suggested by penetration depth \cite{hardy.93}, 
transport\cite{bonn.98,ong.99,chiao.99}, and angle resolved photoemission 
spectroscopy \cite{kaminski.99} (ARPES) experiments. Low temperature 
superconducting (SC) state properties of the cuprates thus appear to
be consistent with d-wave BCS theory with nodal QP excitations. 
However, the importance of
correlations at low $T$ is evident with underdoping: experiments\cite{uemura.89}
show that the superfluid stiffness $D_s(T\!=\!0)\!\sim\!x$, and the QP
weight at $(\pi,0)$ diminishes on approaching the Mott insulator 
\cite{shen-ding.00}. In this paper we address the question of interaction
corrections to the temperature dependence of $D_s(T)$.

The in-plane superfluid stiffness $D_s (T) = 
(c^2 d/4\pi e^2 \lambda^2)$, 
with $d$ the mean interlayer spacing along the
c-axis, can be directly obtained from
measurements of the in-plane penetration depth, $\lambda (T)$. 
$D_s(T)$ is found to
decrease linearly with temperature\cite{hardy.93}
$D_s(T) = D_s(0) - AT$ for $T \ll T_c$, 
with a slope $A$ which is nearly doping independent 
\cite{panagopoulos.98a,lemberger.99}
(or weakly decreasing but nonsingular) as $x\!\to\!0$.

Clearly the linear drop in $D_s(T)$ is due to thermally generated
excitations which contribute to the normal fluid density. 
BCS theory with noninteracting QP excitations around the four
d-wave nodes leads to the result
\be
D_s(T) = D_s(0) - \frac{2 {\rm ln} 2}{\pi}
{{v_{F}}\over{v_{2}}} T.
\label{bcsslope}
\ee
where $v_{F}$ is the Fermi velocity and $v_2$ is related to the slope 
of the SC gap via $v_{2}= (1/k_F) 
\partial\Delta(\theta)/\partial\theta|_{{\rm \theta=\pi/4}}$,
at the nodal Fermi wavevector $k_{F}$. Mesot et al.~\cite{mesot.99} 
obtained the nodal QP dispersion parameters $v_{F}$ and $v_{2}$
as a function of doping from ARPES data on Bi2212, and compared the 
$D_s$ slope obtained from Eq.(\ref{bcsslope}) with $\lambda$ 
measurements. They found that 
the slope estimated in this manner is too large by more than a factor 
of two at optimal doping --- the ARPES results \cite{mesot.99} 
of $v_F\!=\!2.5\times 10^7 {\rm cm/sec}$ and $v_2\!=\!1.25\times 
10^6 {\rm cm/sec}$ lead to an estimated slope $d D_s/dT=0.77 {\rm meV/K}$, 
while the slope obtained from penetration depth experiments \cite{sflee,
jacobs,waldmann} is approximately $0.33 {\rm meV/K}$. Furthermore, 
this discrepancy increases with underdoping since $v_2$ measured
in ARPES decreases marginally leading to a slight {\it increase} in
the estimated slope $d D_s/dT$ on underdoping, while the slope obtained 
from penetration depth experiments in Bi2212 {\it decreases} somewhat with 
underdoping \cite{waldmann}. This is in contrast to the rather striking 
agreement between estimates from thermal transport measurements 
\cite{chiao.99} and ARPES \cite{mesot.99} for the ratio 
$v_F/v_2\!\approx\!20$ at optimal doping in Bi2212.

Following Refs.~\cite{millis.98,durst.00}, we attribute this
discrepancy to residual QP interactions or Fermi liquid corrections. 
We use here a phenomenological superfluid Fermi
liquid theory (SFLT) to explore the effects of lattice anisotropy 
on QP interactions in more detail than in earlier studies;
(see, however ref.~\cite{walker}). 
Some of the results obtained below were summarized without
derivation in a conference report \cite{arun.00a}. 

We note that thermal phase fluctuations \cite{carlson.99} are ignored here, 
since we have shown elsewhere \cite{paramekanti.00}
that a proper treatment of the long-range Coulomb interaction
results in their contribution to $D_s(T)$ being subdominant to that of the 
nodal QPs. 

\smallskip
\noindent{\bf Superfluid Fermi liquid theory:} 
Fermi liquid (FL) theory for a normal Fermi system
is based on the existence of well-defined (coherent) QP
excitations which are adiabatic continuations of the single 
particle excitations of a free Fermi gas. While
transport and ARPES experiments suggest that the normal state of optimal 
and underdoped high-$T_c$ SC's is not a FL,
nevertheless, sharp QP peaks do appear all over 
the Fermi surface (FS) deep in the SC state (for $T\!\!\ll\!\!T_c$). 
Naturally, one is then 
led to consider a description of the SC state and its low lying QP
excitations as an adiabatic continuation of a BCS state with Bogoliubov
QP excitations. 

The approach advocated in refs.~\cite{millis.98,durst.00}, and adopted below, 
assumes that such a SC state may be viewed as a 
correlated FL in which a pairing interaction has been turned on \cite{note.0}. 
In this case, one can use the
superfluid Fermi liquid theory (SFLT) developed many years ago
\cite{larkin.63,leggett.65,leggett.75}, and generalize it to the
anisotropic case. 

For a normal Fermi system, the change in free energy due to a change in the
QP momentum distribution $\dnk$
takes the standard form 
\be \delta F[\dnk] = \sum_{\vk} \xi_\vk^0
\dnk + \frac{1}{2}\sum_{\vk,\vk'} f(\vk,\vk') \dnk \dnkp
\label{landau}
\ee
where $\xi_\vk^0$ is the dispersion for the QP of momentum
$\vk$ in the absence of other QP's, $f(\vk,\vk')$ is the
Landau interaction function, and $\dnk=\sum_{\sigma}\dnks$.
We have ignored the spin-dependent part of $f(\vk,\vk')$ in 
order to simplify the notation; the generalization with spin 
is straightforward, but not relevant for the present discussion.  
We will refer to the QP's obtained by setting $f(\vk,\vk')=0$ in the 
above equation, as non-interacting QP's. The dispersion for these QP's 
is $\xi_\vk^0$ which does include the mass renormalization.

We now use the above functional to calculate the superfluid stiffness at
low temperatures, in two steps: (1) we calculate the diamagnetic
response to a vector potential and (2) we calculate the
renormalization of the current carried by the interacting
QP's, relative to free QP's, and use this to
compute the paramagnetic current correlator of the QP's.  We next use
the above quantities as inputs to a Kubo formula in the QP
basis,
which allows us to determine the superfluid stiffness $D_s(T)$.

\smallskip
\noindent{\bf Diamagnetic term:~}
Let $n^0_\vk$ be the unperturbed equilibrium QP
distribution. In the presence of
the vector potential $n^0_\vk\to n^0_{\vk+e\vA/c}$ leading to
a shift of the momentum distribution $\dnk=n^0_{\vk+e\vA/c}-n^0_\vk$.
We calculate the diamagnetic term \cite{note.1} as the change 
$\delta F$ to order $A^2$:
\bea
\delta F&=& \sum_{\vk}\xi^0_\vk \left(A^\mu\nabla_\mu n^0_\vk +
\frac{1}{2} A^\mu A^\nu \nabla_{\mu\nu} n^0_\vk 
\right) \nonumber \\
&+& \frac{1}{2}\sum_{\vk,\vk'} f(\vk,\vk') A^\mu A^\nu 
\nabla_\mu n^0_\vk \nabla'_\nu n^0_\vk.
\label{dia.0}
\eea
Here we set $e = c=1$,
$\nabla_\mu,\nabla'_\mu$ denote derivatives with respect to
$\vk_\mu$ and $\vk'_\mu$ respectively, where $\mu,\nu=x,y$ and the 
sum over $\mu,\nu$ is implicit.  
The term linear in $\vA$ vanishes, since the
integrand is odd in $\vk$, and we get
\bea
\delta F\!&\!=\!&\! \frac{1}{2}A^\mu A^\nu \left[
\sum_{\vk}\xi^0_\vk \nabla_{\mu,\nu} n^0_\vk 
\!+\! \sum_{\vk,\vk'} f(\vk,\vk')\nabla_\mu n^0_\vk \nabla'_\nu n^0_\vk
\right] \nn \\
&\equiv&\frac{1}{2}A^\mu A^\nu K_{\mu\nu}
\label{dia.1}
\eea
where $K_{\mu\nu}$ is the diamagnetic response.

Given the jump discontinuity in the ``normal'' state, 
QP distribution at the FS, we use 
$\nabla_\mu n^0_\vk = -2 v^0_{\vk\mu}
\delta(\xi_{\vk}^{0})$, where the factor of $2$ arises from summing over 
both spins.
Using the definition $\vv^0_\vk=\vN \xi^0_\vk$ leads to
\bea
K_{\mu\nu}\!\!\!&\!\!=\!\!&\!\! 2 \sum_\vk v^0_{\vk\mu} v^0_{\vk\nu}
\delta(\xi_{\vk}^{0})
\!\!+\!\! 4 \sum_{\vk,\vk'} f(\vk,\vk') v^0_{\vk\mu} v^0_{\vk\nu}
\delta(\xi_{\vk}^{0}) \delta(\xi_{\vk'}^{0}) \nn \\
&\equiv& \alpha_{_F} K^0_{\mu\nu}
\label{dia.2}
\eea
where $K^0_{\mu\nu}\equiv 2 \sum_\vk v^0_{\vk\mu} v^0_{\vk\nu}
\delta(\xi_{\vk}^{0})$ is the diamagnetic term for non-interacting QP's.

\smallskip
\noindent{\bf Quasiparticle current renormalization:~}
The QP energy $\xi_\vk=\xi^0_\vk+\sum_{\vk'} f(\vk,\vk')\dnkp$,
leads to the QP velocity 
$\vv_{\vk}=\vv^0_{\vk}+\sum_{\vk'} \nabla f(\vk,\vk') \dnkp$. The
total QP current $\vJ$ is then $\sum_{\vk} \vv_{\vk} n_{\vk}$, 
which reduces to
\be
\vJ = \sum_{\vk} \vv^0_\vk \dnk - \sum_{\vk,\vk'} f(\vk,\vk') 
\dnkp \vN n^0_\vk,
\label{curr.0}
\ee
where we have used $\sum_{\vk}\vv^0_\vk n^0_\vk=0$ 
in the first term, since the equilibrium QP population does not carry any 
current. In the second term, we have transferred the $\vk$-derivative from
$f(\vk,\vk')$ to $n_\vk$, with $n_\vk\approx n^0_{\vk}$ at this order.
This relates the current carried by the interacting QP, to that carried
by a non-interacting QP which only has a mass renormalization.
%

To make further progress in
the specific case of a d-wave SC, we note that the dominant
excitations in the low temperature state are those near the gap
nodes. We therefore restrict our attention to the renormalization of
the current carried by the QP's at the $4$ nodal points located
at ${\vk}_{_F}^M$, with $M=1\ldots 4$. Setting
$\nabla_\mu n^0_\vk \simeq -2 v^0_{\vk\mu} \delta(\xi_{\vk}^{0})$ as
before, we find that the contribution to the current at the $M$-th
node 
\bea
J_\mu(M)\!&\!\!=\!\!&\! J^0_\mu(M)\left[ 1\!\! +\!\! \frac{2}{v_{_F\mu}(M)}\sum_{\vk'}
f(\vk_{_F}^M, \vk') v^0_{\vk'\mu}
\delta(\xi_{\vk'}^{0}) \right] \nn \\
&\equiv&  J^0_\mu (M) \beta_{_F}
\label{curr.1}
\eea
where $J^0_\mu(M)\equiv v_{_F \mu} (M) \dnk (M)$ is the 
non-interacting QP current.
In arriving at the above result, we have
interchanged the $\vk,\vk'$ labels in the second term, used the
symmetry, $f(\vk,\vk')=f(\vk',\vk)$, and there is no implicit sum 
over $\mu$ in Eq.~(\ref{curr.1}).

\smallskip
\noindent{\bf The superfluid stiffness:} 
From the Kubo formula, we find $D^{\mu\nu}_s =
K_{\mu\nu}-\Lambda_{\mu\nu}(\vq\to 0,i\omega_n=0)$ where
$\Lambda_{\mu\nu}(\vq,i\omega_n)\equiv\la j_\mu(\vq,i\omega_n)
j_\nu(-\vq,-i\omega_n)\ra$ is the current correlator and we take the
transverse limit of $\vq\to 0$. In the QP basis, there are
no excitations at $T=0$ and $D^{\mu\nu}_s(T=0)=K_{\mu\nu}$. At low
temperatures, there are nodal QP excitations and the
current operator in $\Lambda(\vq,i\omega_n)$ has matrix elements
between the ground state and these excited states. The current carried
by the QP's is however renormalized by the factor
$\beta_{_F}$ which leads to $\Lambda=\beta^2_{_F} \Lambda^0$, with
$\Lambda^0$ being the correlator for the non-interacting
QP's. The correlator $\Lambda^0$ is easily evaluated within 
BCS theory using the dispersion $\xi^0_\vk$, and is linear in $T$ at
low temperature in a d-wave SC.  Further, there are polarization
effects by which the flowing QP's lead to an internal (fictitious)
vector potential arising from the $f(\vk,\vk')$, in addition to the
applied vector potential \cite{leggett.75}.  This effect is important
close to $T_c$ when there are a large number of QP's, but it is
unimportant at low temperature when there are very
few thermally excited QP's\cite{leggett.65,leggett.75}.
The superfluid stiffness in a d-wave SC at low $T$ is thus given by
\be
D_s(T) = \alpha_{_F} K^0 - \beta^2_{_F}
\left(\frac{2 \ln 2}{\pi}\frac{v_{F}}{v_2}\right)~T
\label{ds}
\ee
where $K^0=(1/d)~{\rm Tr} K^0_{\mu\nu}$ in $d$-dimensions, assuming cubic
symmetry. We now proceed to discuss the 
FL corrections $\alpha_{_F},\beta_{_F}$ in more detail.

\smallskip
\noindent{\bf Isotropic limit:~}
For an isotropic system $v_{_F}$ and $k_{_F}$ are independent of the 
location on the FS and $m^*\equiv k_{F}/v_{F}$ is the
effective mass. The Landau interaction 
$f(\vk,\vk')\equiv f(\vk\cdot\vk')$ and depends only on the angle
between the two momenta on the FS. 
Retaining only the {\it single} Landau parameter 
relevant for this discussion,
$f(\vk\cdot\vk')=({dn}/{d\epsilon})^{-1} F_1 \cos\theta$, 
where $\cos\theta=\hat{\vk}\cdot \hat{\vk}'$ and 
$(dn/d\epsilon)=m^*/\pi$ is the total ``normal'' state QP density of 
states for both spins. It is then easy to see that in 2D
\be
K_{\mu\nu}=\delta_{\mu\nu} \frac{n}{m^*} (1+F_1/2)
\ee
where the 2D electron density $n = k_{_F}^2/2\pi$. 
From Eq.~(\ref{dia.2}), we thus find $\alpha_{_F}= (1+F_1/2)$.
(For the special case of a Galilean-invariant system, using the Landau 
relation $(1+F_1/2)= m^*/m$ in 2D, we find $K_{\mu\nu}=\delta_{\mu\nu}
(n/m)$).
It is also easy to find that the renormalization of the current in the
isotropic case is given by $J_\mu = J^0_\mu~(1+F_1/2)$, and the 
current correlator is then $\Lambda= \beta^2_{_F} \Lambda^0$ with 
$\beta_{_F} = (1+F_1/2)$.
These results for $\alpha_{_F},\beta_{_F}$ are in agreement with
the earlier work of Larkin and Migdal
\cite{larkin.63} and Leggett \cite{leggett.65}.

We now discuss the shortcomings of isotropic
SFLT as applied to the high $T_c$ SC's following Ref.~\cite{millis.98}.
Low temperature penetration depth experiments \cite{uemura.89} suggest that 
$D_s(x,T=0)\!\sim\!x$. At the same time, ARPES
experiments, as well as theoretical studies of SC
in doped Mott insulators \cite{paramekanti.01},
suggest that $m^*$ does not diverge on underdoping. Within the isotropic
SFLT framework, these two together imply $(1+F_1/2)\!\sim\!x$
which in turn means the slope of $D_s(x,T)$ is 
proportional to $(1+F_1/2)^2\!\sim\!x^2$. This scaling of the slope 
\cite{lee}, 
however, is in strong disagreement with penetration depth measurements. 
Following the suggestion \cite{millis.98} that this problem may be resolved
by including anisotropy of the Landau interaction function
over the FS, we next try to understand FL corrections in the anisotropic case.

\smallskip
\noindent{\bf Anisotropic case:~}
In order to set up a phenomenological SFLT on a 2D square lattice, we 
first rewrite all our functions in terms of an angle variable $\theta$
which sweeps over the large hole-barrel FS centered around $(\pi,\pi)$. 
Then, the Fermi momentum $k_{_F}\equiv k_{_F}(\theta)$,
the Fermi velocity $v_{_F}\equiv v_{_F}(\theta)$ and the Landau
interaction function $f(\vk,\vk')\equiv f(\theta,\theta')$. 
We expand these in an orthogonal basis 
\bea
v_{_{F X}}(\theta) = \sum_{\ell=0}^\infty V^{(\ell)}_{_X} \cos\left[
(2 \ell + 1)\theta\right] \\
v_{_{F Y}}(\theta) = \sum_{\ell=0}^\infty V^{(\ell)}_{_Y} \sin\left[
(2 \ell + 1)\theta\right] \\
k_{_{F}}(\theta) = k_{_{F 0}}+\sum_{\ell=1}^\infty 
k_{_F}^{(\ell)}\cos (4 \ell \theta),
\eea
where we have used the symmetries of the square lattice to restrict the 
form of the expansion, and also used the vector (scalar) character of the 
$v_{_F}$ ($k_{_F}$).
We may also generally expand the interaction, $f(\theta,\theta') = 
\sum_{\ell,m}  F_{\ell,m} e^{i\ell\theta} e^{i m\theta'}$.
We restrict the form of $f(\theta,\theta')$ using the following
symmetries: (i) $f(\theta,\theta')=f(\theta',\theta)$, (ii)
$f(\theta,\theta')=f(-\theta,-\theta')$ and (iii)
$f(\pi/2-\theta,\pi/2-\theta')=f(\theta,\theta')$. While (i) is
generally valid, (ii) and (iii) are valid for a
square lattice. This finally leads to
\be
f(\theta,\theta')\!\! =\!\! \sum_{\ell \geq m}  F_{\ell,m} \left[
\cos(\ell\theta+m\theta') + \cos(\ell\theta'+m \theta) \right]
\ee
where $\ell,m:-\infty\to\infty$ with $(\ell+m)=4 p$ and $p=0,\pm 1,\pm
2,\ldots$. We have set $\ell \geq m$ to avoid overcounting.
We note that: (a) the interaction function depends on
$\theta$ and $\theta'$ separately in general and not only on
$(\theta-\theta')$ as in the isotropic case and (b) there are many
more Landau parameters on the lattice, labeled by two integers
$(\ell,m)$. As we shall see, this considerably complicates our problem
since many Landau parameters may contribute to a given response
function, which prevents their unique determination \cite{swihart}. 
This is unlike the isotropic case (say in He$^3$) where usually a 
single Landau parameter renormalizes a particular correlation function.

We now write the results for $\alpha_{_F},\beta_{_F}$ in these new 
coordinates. The diamagnetic term is given by
\bea
K_{xx}\!\!&\!=\!&\!\! 2 \int_0^{2\pi}\frac{d\theta}{(2\pi)^2} 
\frac{k_{_F}(\theta)}{|v_{_F}(\theta)|} v^2_{_{F X}}(\theta) \\
&\!\!+\!\!&\!\! 4 \int_0^{2\pi}\!\!\frac{d\theta d\theta'}{(2\pi)^4} 
\frac{k_{_F}(\theta)}{|v_{_F}(\theta)|}
\frac{k_{_F}(\theta')}{|v_{_F}(\theta')|}
v_{_{F X}}(\theta) v_{_{F X}}(\theta') f(\theta,\theta')\nn
\eea
and the current renormalization for node-$M$ is
\be
\frac{J_x (M)}{J^0_x (M)}\!\!=\!\!1\!\!+\!\! 2 
\int_0^{2\pi}\!\frac{d\theta'}{(2\pi)^2}
\frac{k_{_F}(\theta')}{|v_{_F}(\theta')|} 
\frac{v_{_{F X}}(\theta')}{v_{_{F X}}(\theta_{_M})}
f(\theta_{_M},\theta')
\ee
where $\theta_{_M}$ is the angular position of node-$M$.
We can express this in a more compact form by defining 
$\la O \ra_{\theta} \equiv 
\int^{2\pi}_0 d\theta k_{_F}(\theta)
O(\theta)/(2\pi|v_{_F}(\theta)|)$.
This yields
\bea
\alpha_{_F}&=&1 + \frac{\la\la v_{_{F X}}(\theta) v_{_{F X}}(\theta')
f(\theta,\theta')\ra\ra_{\theta\theta'}}{\pi \la v^2_{_{F X}}
\ra_{\theta}} \label{alpha.1}\\
\beta_{_F}&=&1 + \frac{\la v_{_{F X}}(\theta') f(\theta_{_M}
,\theta')\ra_{\theta'}}{\pi v_{_{F X}}(\theta_{_M})}
\label{beta.1}
\eea
For $f(\theta,\theta')= (\pi/m^*) F_1 \cos(\theta-\theta')$ 
and $k_{_F},v_{_F}$ 
independent of $\theta$,
we easily recover the isotropic limit.

\smallskip
\noindent{{\bf Simple models for the dispersion and} $f(\vk,\vk')$:~}
We now consider special cases of the general result which serve to 
illustrate the deviation from the isotropic limit.

\smallskip
\noindent{\bf Case I:~} Consider an isotropic dispersion, with $v_{_F}$ 
and $k_{_F}$ independent of $\theta$,
but retain all allowed Landau parameters on the lattice.
In this case, with $m^*\equiv k_{_F}/v_{_F}$, we find
\bea
\alpha_{_F}&=&1+ \frac{m^*}{\pi} (F_{1,1}+F_{1,-1}) \nn \\
\beta_{_F}&=&1 + \frac{m^*}{2\pi} \left(\sum_{p\leq 0} (-1)^p F_{1,4p-1}
+\sum_{p\geq 0} (-1)^p F_{4p+1,-1} \right. \nn\\
&+&\left. \sum_{p > 0} (-1)^p F_{4p-1,1}
+\sum_{p < 0} (-1)^p F_{-1,4p+1}\right)
\eea
Thus, many Landau parameters contribute to the renormalization in this
anisotropic case unlike in the isotropic limit. Furthermore, different 
Landau parameters
contribute to $\alpha_{_F}$ and $\beta_{_F}$. It is then easily possible 
that $\alpha_{_F} \neq \beta_{_F}$ and they could then also behave very 
differently with doping if several Landau parameters are nonzero.

\begin{figure}
\begin{center}
\vskip-2mm
\hspace*{1mm}
\epsfig{file=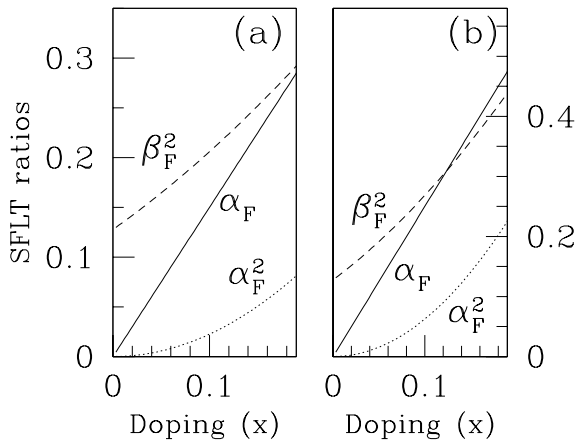,width=8cm,angle=0}
\vskip-2mm
\caption{Doping dependence of the SFLT renormalization ($\beta^2_{_F}$)
of the slope of $D_s(T)$ for a model with anisotropic QP dispersion
and a single Landau parameter chosen such that (a) $\alpha_{_F}(x)=1.5 x$ 
and (b) $\alpha_{_F}(x)=2.5 x$ (see Case II in text for details).
In the isotropic limit, $\beta^2_{_F}(x)=\alpha^2_{_F}(x)$, but there is
marked deviation from this in the anisotropic case --- most strikingly
$\beta^2_{_F}(x)\!\!\ne!\!0$ as $x\!\!\to\!\!0$, as in the
experiments. For this simple model and choice of dispersion, a larger 
renormalization of $D_s(0)$ (smaller $\alpha(x)$ as in panel (a)) appears 
to correlate with a weaker doping dependence of $\beta^2_{_F}(x)$.}
\label{fig.2}
\end{center}
\vskip-5mm
\end{figure}

\smallskip
\noindent{\bf Case II:~}
We next consider the case where we keep a single Landau parameter
$F_{1,-1}\neq 0$, and set all other $F_{\ell,m}=0$. We however retain
the full anisotropy of the dispersion, as measured in ARPES. We
take the tight-binding fit to the (normal state) ARPES 
dispersion \cite{norman.95}, and numerically compute the above
integrals to determine $\alpha_{_F},\beta_{_F}$.
In order to study the doping dependence
of $\alpha_{_F},\beta_{_F}$, we assume a doping dependence $F_{1,-1}(x)= 
B + C x$, such that $\alpha_{_F}(x) \sim x$ in agreement with the
Uemura plot \cite{uemura.89}, with reasonable values $\alpha_{_F}(x=0.2)
\approx 0.3\!-\!0.5$. This 
fixes $B,C$ and we use this to determine the 
doping dependence of $\beta_{_F}(x)$. The result of this calculation
is plotted in Fig.~\ref{fig.2}(a,b) where we see a marked deviation from
the isotropic result ($\beta^2_{_F}=\alpha^2_{_F}$) --- in the 
anisotropic case, $\beta^2_{_F}$ is nonsingular as $x\to 0$, in
qualitative agreement with penetration depth results.

\smallskip
\noindent{\bf Conclusions:~} 
We have used a phenomenological SFLT for a d-wave SC to determine
the renormalization of $D_s(T=0)$ and $d D_s/d T$ due to FL
factors. Within simple models for the dispersion and the Landau
interaction function, we find that anisotropy can cause strong deviations 
from the isotropic result. This allows us to understand the discrepancy 
between penetration depth and photoemission experiments
for the temperature and doping dependence of the superfluid density in
terms of SFLT corrections.
While we discussed the case of a
d-wave order parameter as appropriate for the high $T_c$ SC's,
our results are easily generalized to any 
unconventional SC with point nodes and well-defined QP's.

\smallskip
\noindent{\bf Acknowledgments:~} A.P. was supported through NSF 
grants DMR-9985255 and PHY99-07949, and the Sloan and Packard 
Foundations. M.R. was supported in part through the Swarnajayanti 
Fellowship of the Indian DST.



\begin{thebibliography}{999}

\bibitem {hardy.93} 
W.N.~Hardy, D.A.~Bonn, D.C.~Morgan, R.~Liang, and K.~Zhang,
Phys. Rev. Lett. {\bf 70}, 3999 (1993). 

\bibitem{bonn.98}
A.~Hosseini, R.~Harris, S.~Kamal, P.~Dosanjh, J.~Preston, R.~Liang, 
W.N.~Hardy, and D.A.~Bonn, Phys. Rev. B {\bf 60}, 1349 (1999).

\bibitem{ong.99}
K.~Krishana, N.P.~Ong, Y.~Zhang, and Z.A.~Xu, R.~Gagnon and L.~Taillefer,
Phys. Rev. Lett.{\bf 82}, 5108 (1999). 

\bibitem{chiao.99}
M.~Chiao, R.W.~Hill, C.~Lupien, L.~Taillefer, P.~Lambert, R.~Gagnon 
and P.~Fournier, Phys. Rev. B {\bf 62}, 3554 (2000).

\bibitem{kaminski.99}
A.~Kaminski, J.~Mesot, H.~Fretwell, J.-C.~Campuzano, M.R.~Norman, M.~Randeria,
H.~Ding, T.~Sato, T.~Takahashi, T.~Mochiku, K.~Kadowaki, and H.~Hoechst,
Phys. Rev. Lett. {\bf 84}, 1788 (2000);
A.~Kaminski, M.~Randeria, J.-C.~Campuzano, M.R.~Norman, H.~Fretwell, 
J.~Mesot, T.~Sato, T.~Takahashi, and K.~Kadowaki, Phys. Rev. Lett. 
{\bf 86}, 1070 (2001).


\bibitem{uemura.89}
Y.J.~Uemura, G.M.~Luke, B.J.~Sternlieb, J.H.~Brewer, J.F.~Carolan, W.N.~Hardy, 
R.~Kadono, J.R.~Kempton, R.F.~Kiefl, S.R.~Kreitzman, P.~Mulhern, T.M.~Riseman, 
D.Ll.~Williams, B.X.~Yang, S.~Uchida, H.~Takagi, J.~Gopalakrishnan, 
A.W.~Sleight, M.A.~Subramanian, C.L.~Chien, M.Z.~Cieplak, G.~Xiao, V.Y.~Lee, 
B.W.~Statt, C.E.~Stronach, W.J.~Kossler, and X.H.~Yu, 
Phys. Rev. Lett. {\bf 62}, 2317 (1989).

\bibitem{shen-ding.00}
D.L.~Feng, D.H.~Lu, K.M.~Shen, C.~Kim, H.~Eisaki, A.~Damascelli,
R.~Yoshizaki, J.-I.~Shimoyama, K.~Kishio, G.D.~Gu, S.~Oh, A.~Andrus,
J.~O'Donnell, J.N.~Eckstein, Z.-X.~Shen, Science {\bf 289}, 277 (2000);
H.~Ding, J.R.~Engelbrecht, Z.~Wang, J.-C.~Campuzano, S.-C.~Wang, H.-B.~Yang, 
R.~Rogan, T.~Takahashi, K.~Kadowaki, and D.G.~Hinks, 
Phys. Rev. Lett. {\bf 87}, 227001 (2001).

\bibitem{panagopoulos.98a}
C.~Panagopoulos and T.~Xiang, Phys. Rev. Lett. {\bf 81}, 2336 (1998)
and references therein.

\bibitem{lemberger.99}
B.R.~Boyce, K.M.~Paget and T.R.~Lemberger, cond-mat/9907196.

\bibitem{mesot.99}
J.~Mesot, M.R.~Norman, H.~Ding, M.~Randeria, J.-C.~Campuzano, A.~Paramekanti,
H.M.~Fretwell, A.~Kaminski, T.~Takeuchi, T.~Yokoya, T.~Sato, T.~Takahashi, 
T.~Mochiku, and K.~Kadowaki, Phys. Rev. Lett. {\bf 83}, 840 (1999).

\bibitem{sflee}
S.-F.~Lee, D.C.~Morgan, R.J.~Ormeno, D.~Broun, R.A.~Doyle, J.R.~Waldram 
and K. Kadowaki, Phys. Rev. Lett. {\bf 77}, 735 (1996).

\bibitem{jacobs}
T.~Jacobs, S.~Sridhar, Q.~Li, G.D.~Gu and N.~Koshizuka,
Phys. Rev. Lett. {\bf 75}, 4516 (1995).

\bibitem{waldmann}
O.~Waldmann, F.~Steinmeyer, P.~M\"uller, J.J.~Neumeier, F.X.~R\`egi, 
H.~Savary, and J.~Schneck, Phys. Rev. B {\bf 53} 11825 (1996). This
measurement is restricted to $T > 17K$.


\bibitem{millis.98}
A.J.~Millis, S.M.~Girvin, L.B.~Ioffe, and A.I.~Larkin,
J. Phys. Chem. Solids {\bf 59}, 1742 (1998).

\bibitem{durst.00}
A.C.~Durst and P.A.~Lee, Phys. Rev. B {\bf 62}, 1270 (2000).

\bibitem{walker}
M.B.~Walker, Phys. Rev. B {\bf 64}, 134515 (2001) and cond-mat/0010086
uses a different approach to anisotropic SFLT combining a microscopic 
Hartree-Fock type theory with the Landau functional for QPs. We are 
unable to make direct contact with his results, which differ from ours,
but believe our approach to be more general; see \cite{note.1}.

\bibitem{arun.00a}
A.~Paramekanti and M.~Randeria, Physica C {\bf 341-348}, 827 (2000).

\bibitem{carlson.99}
E.W.~Carlson, S.A.~Kivelson, V.J.~Emery and E.~Man\-ou\-sakis,
Phys. Rev. Lett. {\bf 83}, 612 (1999).

\bibitem{paramekanti.00}
A.~Paramekanti, M.~Randeria, T.V.~Ramakrishnan and S.S.~Mandal,
Phys. Rev. B {\bf 62}, 6786 (2000);
L.~Benfatto, S.~Caprara, C.~Castellani, A.~Paramekanti, and M.~Randeria,
Phys. Rev. B {\bf 63}, 174513 (2001).

\bibitem{note.0}
The question of an underlying ``normal'' FL in the cuprates
at low temperatures is a difficult one. Experimentally, magnetic fields which
destroy SC at low T in some underdoped cuprates seem to lead to an
insulating state. However, fields and temperatures that destroy
SC might well induce fluctuations of the SC order parameter which
destroy coherent QP's.
From a theoretical point of view also the separation of the 
original interaction into a correlation term and a pairing term 
poses issues beyond the scope of this work, and we make no
attempt to address them here.

\bibitem{larkin.63}
A.I.~Larkin and A.B.~Migdal, Sov. Phys. JETP {\bf 17}, 1146 (1963).

\bibitem{leggett.65}
A.J.~Leggett, Phys. Rev. {\bf 140}, A1869 (1965).

\bibitem{leggett.75}
A.J.~Leggett, Rev. Mod. Phys. {\bf 47}, 331 (1975).



\bibitem{note.1}
For electrons with a density-density interaction
the diamagnetic term may be
obtained by shifting $\xi^0_\vk \to \xi^0_{\vk+e\vA/c}$ and computing
the energy to ${\cal O}({\cal A}^2)$.
This is, however, {\it not} valid for QP's, since the
Landau interaction function (in the isotropic limit)
also includes terms of the form $\vk\cdot\vk'$, which resembles
a ``current--current'' interaction. Such effects are apparently
ignored in the approach of ref.~\cite{walker}.

\bibitem{paramekanti.01}
A.~Paramekanti, M.~Randeria, and N.~Trivedi,
Phys. Rev. Lett. {\bf 87}, 217002 (2001).

\bibitem{lee}
The problem with $D_s(0)\!\sim\!x$ and the slope $\sim\!x^2$ in 
disagreement with experiments, has also been pointed out within 
certain microscopic approaches by X.-G.~Wen and P.A.~Lee, Phys. Rev. 
Lett. {\bf 80}, 2193 (1998) and D.-H.~Lee, Phys. Rev. Lett. {\bf 84}, 
2694 (2000).

\bibitem{swihart}
K.~Aoi and J.C.~Swihart, Phys. Rev. B {\bf 7}, 1240 (1973).

\bibitem{norman.95} 
M.R.~Norman, M.~Randeria, H.~Ding and J.-C.~Campuzano, Phys. Rev. B {\bf 52}, 
615 (1995).



\end{thebibliography}
\end{document}